# Harmonic Function Expansion of Nearly Oblate Systems

D. Syer

*Canadian Institute for Theoretical Astrophysics, MacLennan Labs, 60 St. George Street, Toronto M5S 1A7, Ontario.*

**ABSTRACT**

We show how to develop an expansion of nearly oblate systems in terms of a set of potential-density pairs. A harmonic (multipole) structure is imposed on the potential set at infinity, and the density can be made everywhere regular. We concentrate on a set whose zeroth order functions describe the perfect oblate spheroid of de Zeeuw (1985). This set is not bi-orthogonal, but it can be shown to be complete in a weak sense. Poisson's equation can be solved approximately by truncating the expansion of the potential in such a set. A simple example of a potential which is not one of the basis functions is expanded using the symmetric members of the basis set up to fourth order. The basis functions up to first order are reconstructed approximately using 10,000 particles to show that this set could be used as part of an $N$-body code.

## 1 INTRODUCTION

The use of harmonic basis function expansion of spherical or nearly spherical systems has been described by Clutton-Brock (1973), Polyachenko & Shukhman (1981) and, amongst others, Hernquist & Ostriker (1992). Such an expansion is a useful tool for finding approximate solutions to Poisson's equation

$$\nabla^2 \Phi = \rho, \tag{1}$$

where $\rho$ is the density and $\Phi$ the potential. They can be used in simulations of collisionless, and/or collisional particles, including gas (Hernquist & Ostriker 1992, Barnes & Hernquist 1993, Weil & Hernquist 1993) and in normal mode analysis (Robijn 1994). Solutions have the advantage of being smooth and efficient to generate, especially if the expansion can be truncated at low order.

The functions consist of a set of potential-density pairs $\{\Phi_{jlm}, \rho_{jlm}\}$, where $(j,l,m)$ are integers. The potential and density of a general system may be written

$$\Phi(\mathbf{r}) = \sum_{jlm} A_{jlm} \Phi_{jlm}(\mathbf{r}) \tag{2}$$

$$\rho(\mathbf{r}) = \sum_{jlm} A_{jlm} \rho_{jlm}(\mathbf{r}), \tag{3}$$

where

$$\nabla^2 \Phi_{jlm} = \rho_{jlm}, \tag{4}$$

By convention $j$ labels the radial dependence of $\Phi_{jlm}$, and $l$ and $m$ label the polar dependence. In spherical coordinates typically

$$\Phi_{jlm} = F_{jl}(r) P_{lm}(\cos\theta) e^{im\phi}, \tag{5}$$

where $P_{lm}$ is the associated Legendre polynomial. In the works mentioned above $F_{jl}$ can be chosen to make $\{\Phi_{jlm}, \rho_{jlm}\}$ bi-orthogonal, i.e

$$\int \Phi_{ikm} \rho_{jln} d^3\mathbf{r} = \delta_{ij}\delta_{kl}\delta_{mn}, \tag{6}$$

where $\delta_{ij}$ is the Kronecker delta, and hence

$$A_{jlm} = \int \Phi_{jlm}(\mathbf{r}) \rho(\mathbf{r}) d^3\mathbf{r}. \tag{7}$$

Let us write

$$F_{jl}(r) = W_{jl}(r) U_{jl}(r), \tag{8}$$

where $U_{jl} \rightarrow$ constant and $W_{jl}$ has the desired asymptotic form of $\Phi_{jlm}$ as $r \rightarrow 0, \infty$,. We refer to a set in which $W_{jl}$ is chosen to have a multipole structure at infinity, as a 'harmonic set,' *viz.*

$$W_{jl}(r) \sim \frac{1}{r^{l+1}}, \qquad \text{as } r \rightarrow \infty. \tag{9}$$

Clutton-Brock (1973) chose a Plummer model for the zeroth-order potential and

$$W_{jl}(r) = \frac{r^l}{(1+r^2)^{l+1/2}}; \tag{10}$$

Hernquist and Ostriker (1992) chose a Hernquist model and

$$W_{jl} = \frac{r^l}{(1+r)^{l+1}}. \tag{11}$$

The importance of the zeroth-order model lies in the ability to truncate the expansion at low order for systems in which

$\Phi(\mathbf{r}) \simeq \Phi_{000}$.

Saha (1993) points out that considerable flexibility may be gained by dropping the requirement of bi-orthogonality. In this case

$$\int \Phi_{ikm}\rho_{jln} d^3\mathbf{r} = S_{ijklmn}, \tag{12}$$

and the matrix $S_{ijklmn}$ must be inverted to find $\{A_{jlm}\}$. Inversion of the matrix is eqivalent to Gram-Schmidt orthogonalisation of the chosen function set. If the expansion is being truncated at low order, the matrix inversion is not costly, and only has to be done once in any case. The added flexibility of this approach arises in the freedom to specify $W_{00}$ arbitrarily, and thus lower order truncation can be done in systems which are still almost spherical, but are not nearly the same as a Plummer or a Hernquist model.

In this paper we describe a class of function sets that have oblate or disk-like zeroth-order density. We concentrate on a set which has a de Zeeuw (1985) perfect oblate spheroid as the zeroth-order model. We take the view that bi-orthogonality is a luxury, and find the inverse of $S_{ijklmn}$ in low order for the perfect oblate spheroid set. Finally we show that solution of Poisson's equation for a set of 10,000 particles is feasible using the same set.

## 2 A HARMONIC OBLATE SET

We will use a prolate spheroidal polar coordinate system throughout, for which we define the coordinates $(\xi, \eta)$ in the $(R, z)$ plane as follows:

$$z = e\xi\eta, \qquad R^2 = e^2(\xi^2 - 1)(1 - \eta^2). \tag{13}$$

We see that $\xi \in [1, \infty)$ is radius-like, and that $\eta \in [-1, 1]$ is like $\cos\theta$ in spherical polar coordinates. Surfaces of constant $\xi$ are prolate spheroids with foci at $R = 0$, $z = \pm 1$, and they are asymptotically spherical of radius $\xi$ as $\xi \to \infty$. The foci ($\xi = 1$, $\eta = \pm 1$) are singular points of the coordinate system, which we may think of as the end-points of the 'needle' $\xi = 1$.

In these coordinates the Laplacian can be written as

$$e^2\nabla^2 = L(\xi, \eta) + L(\eta, \xi) + \frac{1}{(\xi^2 - 1)(1 - \eta^2)}\frac{\partial^2}{\partial\phi^2} \tag{14}$$

where $\phi$ is the usual polar variable in the $x$-$y$ plane, and

$$L(\xi, \eta) = \frac{1}{\eta^2 - \xi^2}\frac{\partial}{\partial\xi}(1 - \xi^2)\frac{\partial}{\partial\xi}. \tag{15}$$

We recognise in $L(\xi, \eta)$ a close relative of the $\theta$ part of $\nabla^2$ in spherical coordinates. A consequence of this similarity is that

$$Q_{jlm}(\mathbf{r}) = P_{jm}(\xi)P_{lm}(\eta)e^{im\phi} \tag{16}$$

is an eigenfunction of $\nabla^2$, where $P_{lm}$ is the associated Legendre function. Unfortunately the corresponding density is not regular at the points of the needle. The eigenfunctions of $\nabla^2$ which are regular are not expressible in closed form (Abramowicz and Stegun 1964). Thus we are motivated to seek a potential-density set which is regular, but not orthogonal.

One well-known potential in this coordinate system is de Zeeuw's perfect oblate spheroid:

$$\Phi_{\rm pos} = \frac{\xi\tan^{-1}(e\xi) - \eta\tan^{-1}(e\eta)}{\xi^2 - \eta^2}, \tag{17}$$

with the remarkably simple and regular density

$$\rho_{\rm pos} = \frac{2e(1 + e^2)}{(1 + e^2\xi^2)^2(1 + e^2\eta^2)^2} \tag{18}$$

or, in cylindrical coordinates

$$\rho_{\rm pos} = \frac{\rho_0}{(1 + R^2/(1 + e^2) + z^2)^2}, \tag{19}$$

(de Zeeuw 1985, Binney and Tremaine 1987).

As a generalization of (17) let us consider potentials of the form

$$\Phi = \frac{h(\xi, \eta) - h(\eta, \xi)}{\xi^2 - \eta^2}. \tag{20}$$

This, and the associated density, is regular at the points of the needle provided

$$h(\xi, \eta) = h(-\xi, -\eta) \tag{21}$$

(see appendix).

Thus we seek a set of potential functions based on

$$h_{jlm}(\xi, \eta) = W_{jl}(\xi, \eta)\, X_{jm}(\xi)\, Y_{lm}(\eta), \tag{22}$$

where $X_{jm}$ and $Y_{lm}$ combine to give us the 'completeness' that we are looking for in a potential basis set. We demand that $X_{00} = Y_{00} = 1$, so that the function $W_{00}$ determines the zeroth-order potential $\Phi_0$. This should be chosen to match the system we are modelling closely, in which case the function set expansion can be truncated at low order; here we concentrate on

$$W_{00}(\xi, \eta) = \xi\tan^{-1}(e\xi), \tag{23}$$

giving a perfect oblate spheroid. Following Hernquist and Ostriker (1992), we tailor the function $W_{jl}$ to give our set a harmonic structure at infinity. We require that

$$\Phi_{jlm}(\xi, \eta) \sim \frac{1}{\xi^{l+1}}, \qquad \text{as } \xi \to \infty. \tag{24}$$

It will also be useful to require that $W_{jl}$ and at least its first two derivatives are regular everywhere (see appendix).

The choice of the functions $W_{jl}$, $X_{jm}$ and $Y_{lm}$ will be influenced by the requirement that $\rho = \nabla^2\Phi$ is regular. The inclusion of the index $m$ in equation (22) indicates that we intend to extend the $(R, z)$ plane into three dimensions by multiplying $\Phi_{jlm}$ by $e^{im\phi}$. Having done this we find that the $\phi$ part of $\nabla^2$ gives the simplest contribution to $\rho_{jlm}$, *viz.*

$$\frac{1}{(\xi^2 - 1)(1 - \eta^2)}\frac{\partial^2}{\partial\phi^2}\Phi_{jlm} = -\frac{m^2}{(\xi^2 - 1)(1 - \eta^2)}\Phi_{jlm}. \tag{25}$$

Thus we have to be careful about the $z$-axis ($\xi = 1$ and $\eta = \pm 1$), because $\nabla^2$ has a vanishing denominator there

(as in the spherical case). The way we deal with this in the spherical case, where $\cos\theta$ has an $L$-like operator in $\nabla^2$, is to make $\Phi$ proportional to $P_{lm}(\cos\theta)$. In the spheroidal case, both $\xi$ and $\eta$ have the operator $L$ in $\nabla^2$, so we set

$$X_{jm} = P_{jm}, \qquad Y_{lm} = P_{lm}$$

being careful to define $P_{lm}(x)$ in such a way as to make it real for $x > 1$. (Most of the $\{P_{lm}(x)\}$ are not analytic at $x = 1$, so analytic continuation makes no sense, but all we need is a function that $\sim P_{lm}(x)$ as $x \to 1$.) An obvious choice is

$$P_{lm}(x) = a(m)\left|1 - x^2\right|^{|m|} \frac{d^{|m|}}{dx^{|m|}} P_l(x), \tag{26}$$

where $P_l(x)$ is the usual Legendre polynomial and

$$a(m) = \begin{cases} (-1)^m, & |x| < 1 \\ 1, & \text{otherwise.} \end{cases} \tag{27}$$

$a(m)$ is chosen to give the derivatives of $P_{lm}$ the same sign on either side of $x = \pm 1$. The appendix shows that this choice of $X_{jm}$ and $Y_{lm}$ gives a regular density on the $z$-axis.

Having chosen $X_{jm}$ and $Y_{lm}$, we turn to $W_{jl}$ and to the points of the needle. We know that $P_{lm}(x) = (-1)^{l+m}P_{lm}(-x)$, and $h_0(\xi,\eta)$ has been chosen to be symmetric, so to ensure the regularity of $\rho_{jlm}$ at the points of the needle we require that

$$W_{jl}(\xi,\eta) = (-1)^{j+l} W_{jl}(-\xi,-\eta). \tag{28}$$

Assuming that $W_{00}(\xi,\eta)$ is symmetric, an example would be

$$W_{jl}(\xi,\eta) = \frac{\xi^{j+l}}{(1 + e^2(\xi^2+\eta^2))^{j+l}} W_{00}(\xi,\eta). \tag{29}$$

Putting all the above together, a well-behaved set whose zeroth order potential is the perfect oblate spheroid is given by

$$h_{jlm}(\xi,\eta) = \frac{\xi^{j+l}}{(1 + e^2(\xi^2+\eta^2))^{j+l}}\, \xi \tan^{-1}(e\xi)\; P_{jm}(\xi) P_{lm}(\eta). \tag{30}$$

Other sets with regular density can be trivially derived from this by replacing $\xi\tan^{-1}(e\xi)$ with some other symmetric function of $\xi$ and/or $\eta$.

Unfortunately, none but the zeroth-order density function (18) can be expressed in compact form.

## 3 RESULTS

The four lowest order density functions of the perfect oblate spheroid set are shown in Figures 1 and 2. The zeroth- order density is smooth and everywhere positive. The higher order functions are more oscillatory as one would expect, and are sometimes negative. All the density functions have finite or zero mass since they are regular and have a multipole structure at infinity.

Table 1 lists the matrix elements of $S_{ijklmn}$ in a square grid for an eccentricity $e$ of 0.5. Because of the orthogonality of the perfect oblate spheroid function set in the $\phi$ direction, we only need consider $m = n$; and since $\Phi_{jlm} = 0$ for $m > \min[j,l]$ we can ignore larger values of $m$. The map between the new index $p$ and $(jlm)$ is defined in the table. The resulting matrix is still quite sparse because $S_{ijklmn} = 0$ if $k + l$ is odd, by symmetry. The integrations were all carried out over a finite volume bounded by the curve $\xi = 3e$. All elements are accurate to approximately 10 significant figures.

Consider the potential-density pair $(\Phi, \rho)$ in which

$$\Phi(\mathbf{r}) = \Phi_{000}(e_1,\mathbf{r}) + a\Phi_{000}(e_2,\mathbf{r}), \tag{31}$$

where $e_1, e_2$ are eccentricities, and $a$ is a constant. We can try to approximate $\Phi(\mathbf{r})$, for instance, by expanding in the function set $\{\Phi_{jlm}(e_1,\mathbf{r})\}$. Thus we construct the vector

$$b_{ikm} = \int \Phi_{ikm}(e_1,\mathbf{r})\rho(\mathbf{r}), \tag{32}$$

and find $\{A_{ikm}\}$ by solving the linear equations

$$\sum_{jln} S_{ijklmn} A_{jln} = b_{ikm}. \tag{33}$$

The result of doing this in the case $(e_1, e_2) = (0.5, 1.0)$, $a = 0.1$ is shown in Figure 3. The values of $A_{jlm}$ were found for the symmetric ($l$ even, $m = 0$) functions up to order $(j,l) = (4,4)$ (15 functions in all). The solid curve shows the density $\rho$ along the $z$-axis. The lower dashed curve is the zeroth order approximation; the upper dashed curve is the expansion using the values $(jlm) = \{000, 020, 040, 100\}$; and the expansion using all 15 functions is indistinguishable from the solid curve.

Consider a swarm of particles with positions $\{\mathbf{r}_a\}$ drawn from a distribution proportional to $|\rho_q|$, and with masses $\{m_a\}$, all equal in magnitude, but negative wherever the density is negative. The quantity

$$b_p = \sum_a m_a \Phi_p = \hat{S}_{pq}, \tag{34}$$

should be approximately equal to the matrix element $S_{pq}$ and the quantity

$$\hat{\delta}_{pq} = \sum_r S^{-1}_{pr} \hat{S}_{rq}, \tag{35}$$

should be approximately equal to the Kronecker delta. In Table 2 we show the results of testing this assertion for the lowest five density functions using 10,000 particles. It shows that $\hat{\delta}_{pq}$ is approximately equal to $\delta_{pq}$, at least within the expected tolerance, given that statistical errors are expected to be a few percent. We might think of this as a minimum requirement of feasibility for the function set to be used in particle simulations.

## 4 DISCUSSION

An aesthetic drawback of using function sets which are not bi-orthogonal is that we are sometimes not sure quite how complete they are, and we hesitate to call our set a 'basis set' until we know what space it spans. Of course one can always make the tautological statement that $\{\Phi_{jlm}\}$ is a complete basis of the subspace of all functions spanned by $\{\Phi_{jlm}\}$. The question is then, is this subspace a useful one; can it represent a sufficiently broad class of sensible potential-density pairs for practical purposes? The spherical function sets referred to in Section 1 are the eigenfunctions of a hermitian operator, hence they are bi-orthogonal, and they form a complete bi-normal set (i.e. any function for which $\Phi\nabla^2\Phi$ is integrable can be represented by a linear combination of the set). For practical purposes formal completeness is not an issue of great urgency, for one can only ever use a finite subset of even a rigorously complete basis. The best one can do might be, as in the present work, to present a set of functions that closely resembles some zeroth order model of interest, in the hope of finding the smallest possible subspace which is practically useful. We present some arguments below which justify the choice of $h_{jlm}$ in equation (30) by appealing to a rather weak formal completeness.

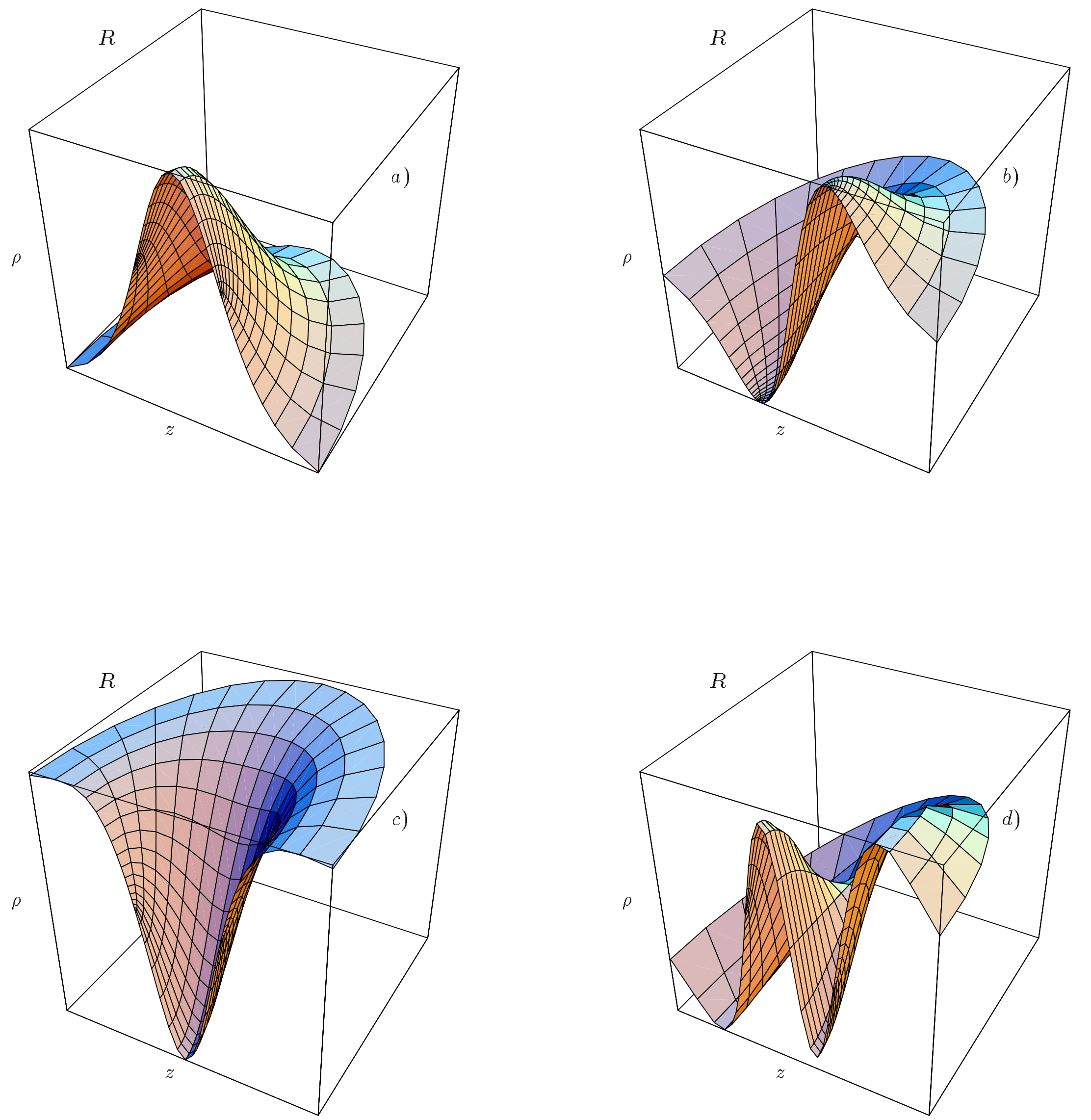

**Figure 1.** The four lowest order densities of the set defined by equation (30) with $e = 0.5$ in the $(x, z)$ plane: a) $(j, l, m) = (0, 0, 0)$ is the perfect oblate ellipsoid; b) $(0, 1, 0)$; c) $(1, 0, 0)$; and d) $(1, 1, 0)$.

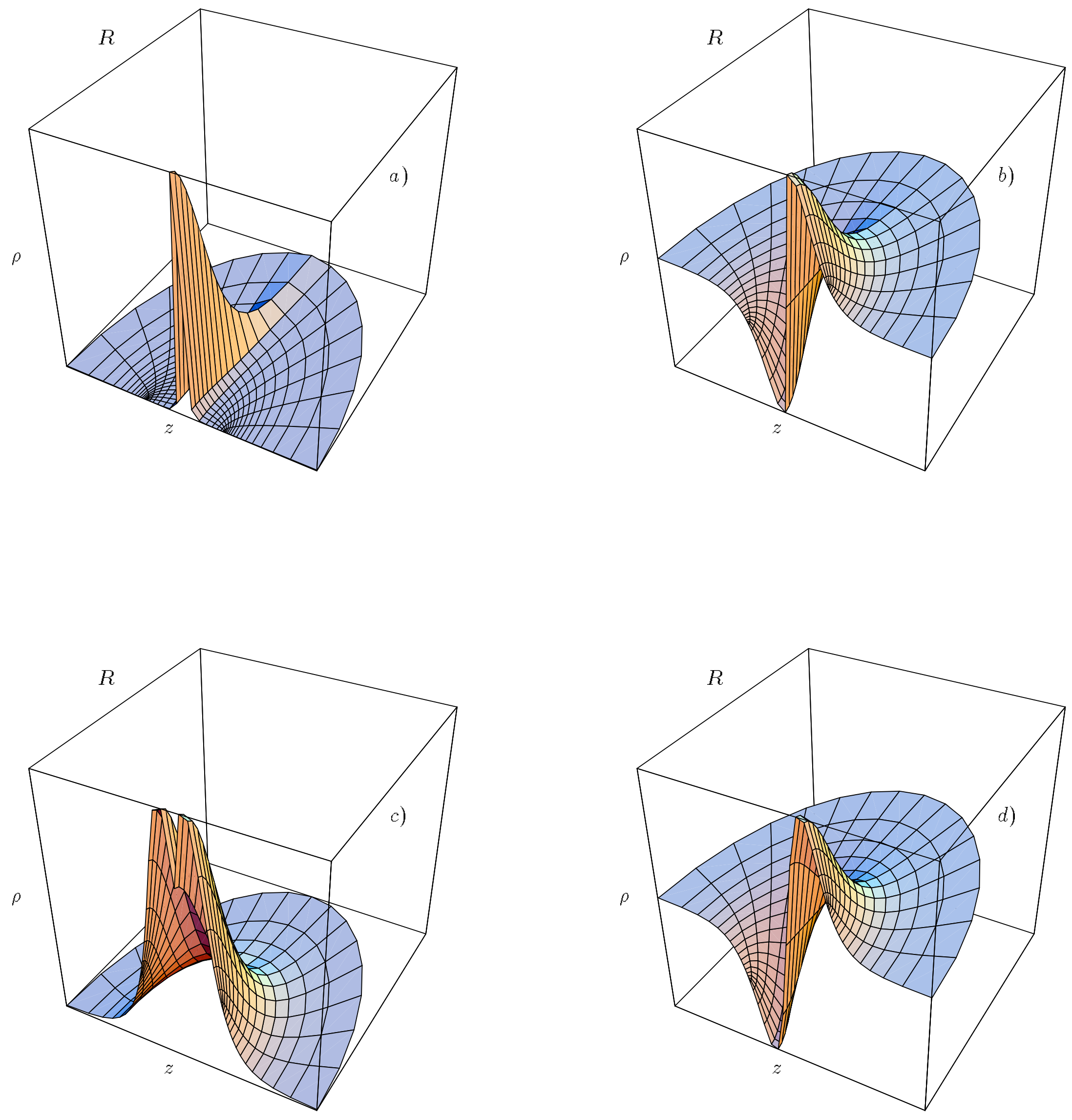

**Figure 2.** As Figure 1 with $e = 10$ in the $(x, z)$ plane: a) $(j, l, m) = (0, 0, 0)$ is the perfect oblate ellipsoid; b) $(0, 1, 0)$; c) $(1, 0, 0)$; and d) $(1, 1, 0)$.

Saha (1993) considers sets of the form (8), but in which $W_{jl}$ does not depend on $j$ (as is the case in all the examples mentioned). Following him, we consider the completeness of such sets from the point of view of approximation in the mean (Courant and Hilbert 1937, Chapter II). Consider the radial part of the $l$th harmonic, and denote the error in the radial part of the potential by $\delta F_l(r)$. If $U_{jl}(r)$ is a polynomial in some variable $u(r)$ with finite range and domain, then it can be shown that the average error

$$\int \frac{|\delta F_l|^2}{W_l^2} \frac{\mathrm{d}u(r)}{\mathrm{d}r} \,\mathrm{d}r \tag{36}$$

may be made arbitrarily small by increasing the order of the polynomial (i.e. by truncating the series at larger $j$). This proves completeness of $\{U_{jl}\}$ over the space of square integrable, piecewise continuous functions of $u$ (*ibid.*). Thus $\{U_{jl}\}$ is complete in a very definite sense, but not neces-sarily $\{F_{jl}\}$. For a finite mass distribution, the $\{F_{jl}\}$ are necessarily harmonic and so completeness of $\{U_{jl}\}$ should be sufficient for practical purposes.

**Table 1.** The matrix $S$ made square and indexed according to $p$. Each element is accurate to approximately 10 significant figures

| $(jlm)$ | $p$ | 1 | 2 | 3 | 4 | 5 |
|---|---|---|---|---|---|---|
| (000) | 1 | 0.45815368137 | 0 | 0.89588371437 | 0 | 0 |
| (100) | 2 | 0 | 0.5186873810 | 0 | 1.283914361 | 0 |
| (010) | 3 | 1.162423755 | 0 | 3.626567178 | 0 | 0 |
| (110) | 4 | 0 | 1.313280882 | 0 | 3.754218875 | 0 |
| (111) | 5 | 0 | 0 | 0 | 0 | 3.677108173 |

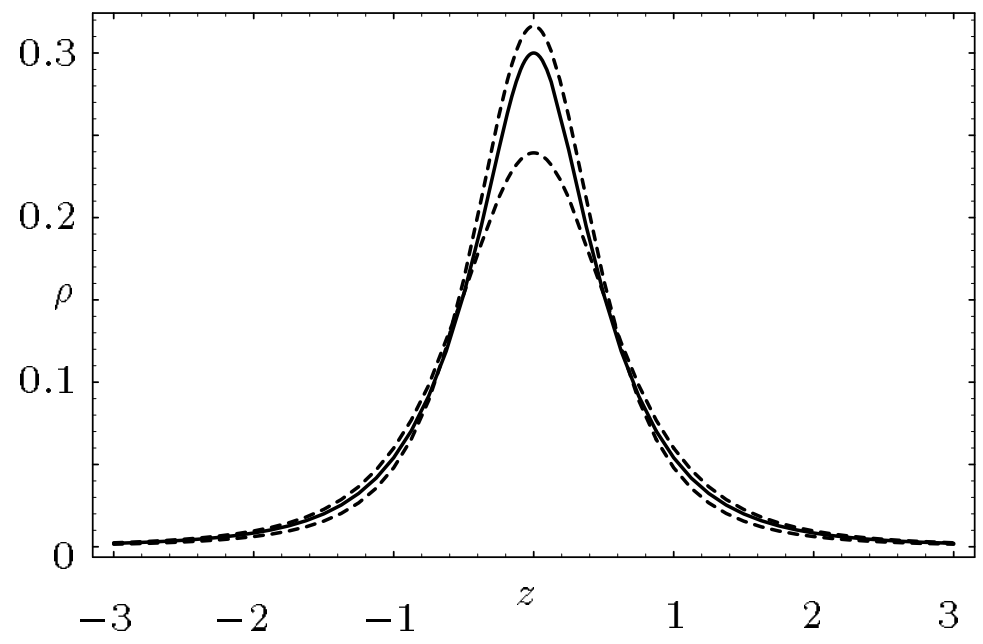

**Figure 3.** The density along the $z$-axis of a simple model (solid line), and successive approximations to it using the perfect oblate spheroid set (dashed lines).

**Table 2.** The matrix $\hat{S}$ calculated using 10,000 particles.

| $(jlm)$ | $p$ | 1 | 2 | 3 | 4 | 5 |
|---|---|---|---|---|---|---|
| (000) | 1 | 1.005 | −0.011 | −0.001 | 0.004 | 0.0003 |
| (100) | 2 | −0.019 | 1.029 | −0.0005 | −0.011 | −0.003 |
| (010) | 3 | −0.050 | 0.042 | 1.01 | −0.010 | −0.005 |
| (110) | 4 | −0.018 | −0.045 | −0.003 | 1.019 | 0.006 |
| (111) | 5 | −0.057 | −0.013 | 0.006 | −0.005 | 1.01 |

Notice the denominator in (29) above. It was chosen so that in the spherical limit ($e \to 0$), $\Phi_{jlm}$ tends to a set which is complete in the sense just described. The spherical limit is also approached at large radius by sets with finite $e$. Specifically, in this limit,

$$\Phi_{jlm} = \frac{r^{l-1}}{(1+r^2)^l} \tan^{-1}(r) \left(\frac{r^2}{1+r^2}\right)^j P_{lm}(\cos\theta)\, e^{im\phi}. \quad (37)$$

This is in the form of (5) with $F_{jl}$ in the form of (8), and hence the weak completeness is ensured.

In fact, we can go a stage further: writing $W_{jl}$ in the form of Clutton-Brock (1973) we would extract

$$U_{jl}(r) = (1+r^2)^{1/2} \frac{\tan^{-1}(r)}{r} \left(\frac{r^2}{1+r^2}\right)^j. \quad (38)$$

We denote the corresponding sets of Clutton-Brock by $\{F_{\rm CB}\}$ and $\{U_{\rm CB}\}$. The $\{U_{\rm CB}\}$ turn out to be a particular set of ultraspherical, or Gegenbauer, polynomials. Clearly the $\{U_{jl}\}$ are of finite range and their domain is the same as the $\{U_{\rm CB}\}$. The two sets are both linearly independent and for every member of $\{U_{jl}\}$ there is a member of $\{U_{\rm CB}\}$. This is no more than a necessary condition for completeness of $\{U_{jl}\}$, but completeness of $\{U_{jl}\}$ would imply completeness of $\{F_{jl}\}$ because $\{F_{\rm CB}\}$ is complete. This gives us some confidence that $\{F_{jl}\}$ might be of practical use.

A simpler looking, but less successful choice of $\Phi_{jlm}$ would be as the above, with $(\xi^2+\eta^2)$ replacing the quantity in brackets in the denominator of $h$. This is similarly well-behaved in terms of its density, but in the limit $e \to 0$

$$\Phi_{jlm} = \frac{r^{l-1}}{(1+r^2)^l} \tan^{-1}(r)\, P_{lm}(\cos\theta)\, e^{im\phi}, \quad (39)$$

in which we see that $\{\Phi_{jlm}\}$ are not even linearly independent (they do not depend on $j$), and the corresponding set of potential-density pairs thus falls far short of being complete. (It could only be used to model systems with a very restricted radial dependence.)

In the spherical case the requirement that the potential is harmonic at leads to a set in which each member is bi-normal and has finite mass. For the works mentioned in the Section 1 it also leads to a bi-orthogonal set. For the present purposes we would admit that the harmonic sructure is not strictly necessary, but it has features which may be considered attractive. For example, each member of the potential set is again automatically bi-normal and has finite mass. Also, members of the density set with $l \neq 0$ all have zero mass as a consequence of their formal similarity to spherical multipoles at infinity. Thus, all the mass in a given distribution resides in the $l = 0$ part of the expansion. Also, the higher order parts of the expansion bear a familiar relationship to their $l = 0$ counterparts. Thus, for instance, our expectation that an $l = 1$ mode should be 'dumb bell' shaped, or dipolar, is confirmed in Figure 1.

It is evident from Figure 2 that the members of the perfect oblate spheroid set with $l \neq 0$ have a density which is considerably less flattened than those with $l = 0$. This problem is probably generic to the coordinate system, which becomes progressively less similar in shape to the zeroth order model as $e$ increases. The perfect oblate ellipsoid set is unlikely to be practically useful in the limit of very flat systems. The density in all the members falls off as a power-law in radius. Thus they are also unlikely to be successful at reproducing realistic disc models, which have exponential decay of the density in the $z$-direction.

Many elliptical galaxies are thought to have density cusps in their inner parts. The spherical bi-orthogonal set of Hernquist and Ostriker (1992) is particularly successful at reproducing cuspy density distributions owing to the presence of a singularity in the density set at the origin, associated with the coordinate singularity there. All of the perfect oblate spheroid density set are smooth and continuous, and so are unlikely to be successful at reproducing cuspy density distributions.

A mass distribution that has constant density $\rho(\tau^2)$ on the self similar ellipsoids

$$\text{constant} = \tau^2 = \frac{x^2}{a^2} + \frac{y^2}{b^2} + \frac{z^2}{c^2}, \tag{40}$$

has a potential given by

$$\Phi = \int_0^\infty \frac{\mathrm{d}u}{\Delta(u)} \int_{\tau^2(u)}^\infty \rho(t^2)\mathrm{d}t^2, \tag{41}$$

where

$$\tau^2(u) = \frac{x^2}{a^2+u} + \frac{y^2}{b^2+u} + \frac{z^2}{c^2+u}, \tag{42}$$

and

$$\Delta(u) = \sqrt{(a^2+u)(b^2+u)(c^2+u)} \tag{43}$$

(Chandresekhar 1969, see also de Zeeuw 1985). So an attempt at constructing a basis set in ellipsoidal coordinates with a cusp at the origin could be made, along similar lines to the Hernquist model, by setting

$$\rho(\tau^2) = \frac{1}{\tau}\frac{\mathrm{d}}{\mathrm{d}\tau}\frac{f(\tau)}{1+\tau^2}. \tag{44}$$

Then

$$\Phi = \int_0^\infty \frac{\Delta(u)}{(\lambda+u)(\mu+u)(\nu+u)} f(\tau(u))\,\mathrm{d}u. \tag{45}$$

(The perfect ellipsoid appears as the case $f(\tau) = \text{constant}$.) However, it is difficult to choose $f(\tau)$ in such a way as to make (45) integrable, while retaining the singularity in $\rho$.

Finally, we note that the a triaxial set of potential density pairs could be constructed by a very similar method. Consider the ellipsoidal coordinate system $(\lambda,\mu,\nu)$, where the coordinates are the roots for $u$ of the equation $\tau^2(u) =$ const (see for example de Zeeuw 1985). The coordinate system defined by equation (13) is the case

$$\frac{a^2}{c^2} = \frac{b^2}{c^2} = 1 + e^2.$$

The perfect triaxial ellipsoid in these coordinates has a potential of the form

$$\begin{aligned}\Phi_{pte} =& \frac{F(\lambda)}{(\lambda-\mu)(\lambda-\nu)} + \\ & \frac{F(\mu)}{(\mu-\nu)(\mu-\lambda)} + \frac{F(\nu)}{(\nu-\lambda)(\nu-\mu)},\end{aligned} \tag{46}$$

so, by analogy with equation (20), we would be lead to consider a potential set of the form

$$\begin{aligned}\Phi =& \frac{h(\lambda,\mu,\nu)}{(\lambda-\mu)(\lambda-\nu)} + \\ & \frac{h(\mu,\lambda,\nu)}{(\mu-\nu)(\mu-\lambda)} + \frac{h(\nu,\lambda,\mu)}{(\nu-\lambda)(\nu-\mu)}.\end{aligned} \tag{47}$$

This form guarantees regularity of the potential at the singular points of the coordinate system.

## 5 APPENDIX

In this appendix we will show first the condition on $h(\xi,\eta)$ which makes $\Phi$ regular at the points of the needle $(\xi,\eta) = (1,-1)$. Let us expand $\nabla^2\Phi$ about the points $(\xi,\eta) = (1,\pm1)$ in a Taylor series in $\eta$. In fact, regularity is trivially satisfied at $\eta = +1$, so we only give the result for $\eta = -1$:

$$\Phi(1,-1) = \frac{h(1,-1) - h(-1,1)}{2(1-y)} + O(1-y)^0. \tag{A1}$$

Thus $\Phi$ is regular at $(\xi,\eta) = (1,-1)$ provided

$$h(1,-1) = h(-1,1). \tag{A2}$$

Next let us expand $\rho = \nabla^2\Phi$ about $(\xi,\eta) = (1,-1)$:

$$\begin{aligned}\rho(1,-1) =& \frac{H}{2(1-y)^2} \\ &+ \frac{3H + 2\,(h_{11}(1,-1) - h_{11}(-1,1))}{4(1-y)} + O(1-y)^0,\end{aligned} \tag{A3}$$

where

$$\begin{aligned}H =& G + h_{10}(-1,1) + h_{10}(1,-1) \\ &- h_{01}(-1,1) + h_{01}(1,-1),\end{aligned} \tag{A4}$$

and

$$G = h(1,-1) - h(-1,1). \tag{A5}$$

In addition to (A2) then, the extra conditions for $\rho$ to be regular are

$$h_{10}(-1,1) + h_{10}(1,-1) - h_{01}(-1,1) + h_{01}(1,-1) = 0, \tag{A6}$$

$$h_{11}(1,-1) - h_{11}(-1,1) = 0. \tag{A7}$$

To satisfy all the regularity conditions (A2), (A6) and (A7) it is sufficient that $h(\xi,\eta) = h(-\xi,-\eta)$.

Next we show that $X_{jm} = Y_{jm} = P_{jm}$ is sufficient to make $\rho$ regular on the $z$-axis, $\xi = 1, \eta < 1$ and $\xi > 1, \eta = \pm1$. First let us define

$$L_\xi = \frac{\partial}{\partial\xi}(\xi^2-1)\frac{\partial}{\partial\xi}. \tag{A8}$$

And then we write

$$h(\xi,\eta) = W(\xi,\eta)\,P_jm(\xi)\,P_lm(\eta), \tag{A9}$$

where at least the first two derivatives of $W(\xi,\eta)$ are regular, and notice that

$$L_\xi P_{jm}(\xi) = \frac{m^2}{(\xi^2-1)}Q_{jm}(\xi) + \langle\text{regular stuff}\rangle\,. \tag{A10}$$

Now

$$\nabla^2\Phi = \frac{1}{\xi^2-\eta^2}\left(L_\xi - L_\eta\right)\Phi - \frac{m^2}{(\xi^2-1)(1-\eta^2)}\Phi,$$

which we can write as

$$\begin{aligned}
&\nabla^2\Phi = \\
&\left\{\frac{1}{\xi^2-\eta^2}\left(\frac{1}{\xi^2-1}+\frac{1}{1-\eta^2}\right)\right. \\
&\left. -\frac{1}{(\xi^2-1)(1-\eta^2)}\right\} m^2\Phi \\
&+\langle\text{regular stuff}\rangle\,.
\end{aligned} \tag{A11}$$

The quantity in braces vanishes, and hence $\rho$ is regular.